\documentclass[runningheads,draftcopy]{svmult}

\usepackage{makeidx}   % allows index generation
\usepackage{graphicx}  % standard LaTeX graphics tool
\usepackage{subeqnar}  % subnumbers individual equations
\usepackage{multicol}  % used for the two-column index
\usepackage{physprbb}  % modified textarea for proceedings,
\makeindex             % used for the subject index
\newcommand{\eq}[1]{{\frenchspacing Eq.~(\ref{#1})}}
\newcommand{\fig}[1]{{\frenchspacing Fig.~(\ref{#1})}}

\begin{document}
\title*{Computing the $\eta$ and $\eta'$ 
Mesons in Lattice QCD}
\toctitle{The Computation of the $\eta$ and $\eta'$ 
Mesons\newline 
in  Lattice QCD}
% allows explicit linebreak for the table of content
%
%
\titlerunning{The Computation of  $\eta$ and $\eta'$ 
in LQCD}
% allows abbreviation of title, if the full title is too long
% to fit in the running head
%
\author{Klaus Schilling%\inst{1}
\and Hartmut Neff\inst{2}
\and Thomas Lippert\inst{1}
}
\authorrunning{K. Schilling et al. }
% if there are more than two authors,
% please abbreviate author list for running head
%
%
\institute{Wuppertal  University, D42097 Wuppertal, Germany
\and Department of Physics, Boston University, Boston MA02215, USA
}  
 \maketitle              % typesets the title of the contribution

\begin{abstract}
  It has been known for a long time that the large experimental singlet-octet
  mass gap in the pseudoscalar meson mass spectrum originates from the anomaly
  of the axial vector current, i.e. from nonperturbative effects and the
  nontrivial topological structure of the QCD vacuum.  In the
  $N_{colour}\rightarrow \infty$ limit of the theory, this connection
  elucidates in the famous Witten-Veneziano relation between the $\eta'$-mass
  and the topological susceptibility of the quenched QCD vacuum.  While
  lattice quantum chromodynamics (LQCD) has by now produced impressive high
  precision results on the flavour nonsinglet hadron spectrum, the
  determination of the pseudoscalar singlet mesons from direct correlator
  studies is markedly lagging behind, due to the computational complexity in
  handling observables that include OZI-rule violating diagrams, like the
  $\eta'$ propagator. In this article we report on some recent progress in
  dealing with this numerical bottleneck problem.

\end{abstract}

\section{Introduction}
\begin{figure}[t]
\begin{center}
\includegraphics[width=.5\textwidth]{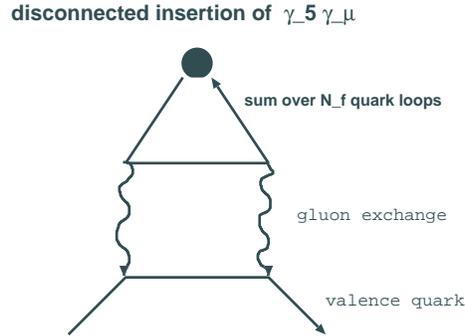}
\end{center}
\caption[]{Perturbative triangle diagram of flavour singlet axial current 
as disconnected diagram with $\gamma_5\gamma_{\mu}$ insertion.}
\label{Fig:pic_axial}
\end{figure}

\begin{figure}[t]
\begin{center}
\includegraphics[width=.5\textwidth]{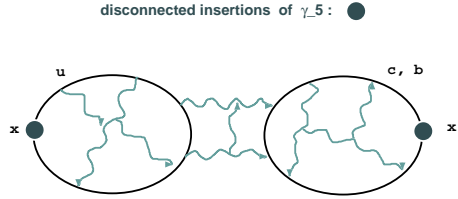}
\end{center}
\caption[]{OZI rule violating quark fusion process with disconnected quark loops and $\gamma_5$ insertions.}
\label{Fig:OZI}
\end{figure}

Long before the advent  of QCD as the field theory of strong
interactions, the $\pi$-meson has been allocated the r\^ole of the
Goldstone Boson in a near-chiral world of light quarks while its
flavour singlet partner, the $\eta'$ meson, was thought to acquire its
nearly baryonic mass of $960$ MeV through ultraviolet quantum
fluctuations that prevent the flavour singlet axial vector current

\begin{equation}
\label{Eq:ax_curr}
j_5^{\mu (0)}(x) = \bar{q}(x)\gamma_5\gamma^{\mu}q(x)
\end{equation}
from being conserved, even in the massless theory.  In this scenario the
breaking of $SU(3)_L \times SU(3)_R $ chiral symmetry occurs as a
renormalization effect within a Wilson expansion of the operator
$\partial_{\mu}j_5^{\mu (0)}$ that suffers operator mixing with the
topological charge density 
\begin{equation}
Q(x) = (g^2/32\pi^2)\; \mbox{tr} F(x)^{\alpha \beta}F^D_{\alpha \beta}(x)
\label{Eq:topol_charge}
\end{equation}
in the form
\begin{equation}
\label{Eq:ax_anom}
 \partial_{\mu}j_5^{\mu (0)}(x)   = 2N_f Q(x) :=  \tilde{Q}(x)\; \; .
\end{equation}
The pseudoscalar operator on the right hand side of this ABJ
anomaly~\cite{ABJ} equation is proportional to the number of active quark
flavours, $N_f$ and contains the gluonic field tensor, $F$, and its dual,
$F^D$.  Its $N_f$ dependence is indicative of ``disconnected'' quark loop
contributions as they appear in the socalled triangle diagrams of perturbation
theory, see \fig{Fig:pic_axial}.  Disconnected diagrams are peculiar to matrix
elements of flavour singlet operators and can be characterized by intermediate
states free of quarks, i.e. by quark-antiquark annihilation into gluons with
subsequent pair creation processes. Their phenomenological r\^ole has been
studied e.g.  in various strong interaction fusion processes where light quark
antiquark pairs annihilate strongly into charm (or bottom) quark antiquark
pairs:
\begin{equation}
\label{Eq:ozi}
u + \bar{u} \rightarrow c + \bar{c}\quad \mbox{or} \quad   b  + \bar{b}\; .
\end{equation}
These  processes, as depicted in \fig{Fig:OZI}, are said to violate the
OZI-rule~\cite{OZI}.

Inspection of the renormalization procedure for this triangle diagram reveals
a clash between quantum physics and symmetry requirements: the renormalizazion
program poses the alternative of either giving up gauge invariance or chiral
symmetry. Accordingly, \eq{Eq:ax_anom} reflects nonconservation of chiral
symmetry, for the sake of preserving gauge invariance, in form of the
ABJ-anomaly (for an excellent review on the subject, see the Schladming
lectures of Crewther~\cite{Crewther}).

%FIG1

The lesson to be learnt is that pseudoscalar flavour singlet mesons are deeply
affected by and thus indicative for the topological structure of the QCD
vacuum.  Needless to say that the time averaged vacuum expectation value of
$Q= \tilde{Q}/(2N_f)$ itself is zero, $\langle 0|Q(\vec{x})|0\rangle =0$, by
translation invariance and parity conservation of strong
interactions\footnote{There is no spontaneous breaking of parity in strong
  interaction physics.}.  As a consequence, the net topological charge of the
QCD vacuum state is zero. The key quantity describing topological fluctuations
of the QCD vacuum is the topological susceptibility, $\chi$, which is defined
by
\begin{equation}
\label{Eq:suscep}
 \chi = \int d^4x \langle 0|\, Q(x)Q(0)\, |0\rangle \;.
\end{equation}

In the so-called 't Hooft limit of the theory, when the
number of colours, $N_c$ is taken to infinity ($N_{c}\rightarrow
\infty$ at  fixed value for the number of active flavours, $N_f$, in
the weak gauge coupling limit $g^2 \sim 1/N_c$), Witten and Veneziano
were able to derive a simple relation for the $\eta'$ mass
in the chiral limit~\cite{WV}
\begin{equation}
\label{Eq:WW}
m^2_{\eta'} = \lim_{N_c \to \infty} \frac{2N_f}{F_{\pi}^2}\chi|_{quenched}\; ,
\end{equation}
where $F_{\pi}$ denotes the pion decay constant.  Within the Witten-Veneziano
model assumption, that the real world is well described by the above limit
relation, LQCD can predict the $\eta'$ mass by computing the topological
susceptibility in quenched QCD. 

The lattice implications of \eq{Eq:WW} have been discussed in quite some
detail in a recent paper of GIUSTI et al~\cite{lattice_WV}. In this
contribution we will focus on those problems that one encounters in the {\it
  direct} approach to extract physics from fermionic two-point correlators in
the singlet sector.

%FIG2

\section{Prolegomena: pseudoscalars in LQCD}
In this section we shall consider a symmetric world with three flavours to
enter the discussion on  the computational tasks to be met.
\subsection{Flavour octet sector}
\label{Sec:octet}
\begin{figure}[t]
\vskip 1cm
\begin{center}
\includegraphics[width=.6\textwidth]{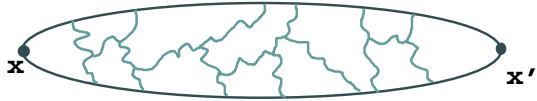}
\end{center}
\vskip -2cm
\caption[]{The connected  propagator $G^{(8)}_{\mbox{PS}}(x,x')$ 
  for  pseudoscalar meson states with negative charge, with a quark $d$ and an
  antiquark $\bar{u}$ running from source $x$ to sink $x'$ in the unquenched
  QCD background field, with  $\gamma_5$ insertions.}
\label{Fig:g8}
\end{figure}

\begin{figure}[ht]
\begin{center}
\includegraphics[width=.5\textwidth]{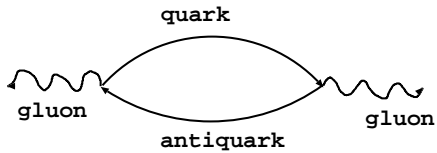}
\end{center}
\caption[]{Gluon propagators including quark loop self energy diagrams.}
\label{Fig:pola}
\end{figure}

In the flavour octet sector of QCD (with mass degenerate $u, d, s$ quarks),
the nonconservation of the axial vector current is determined by the PCAC
relation:
\begin{equation}
\label{Eq:pcac}
 \partial_{\mu}j_5^{\mu (8)}(x)  =  2m \bar{q}\gamma_5q \;,
\end{equation}
which is much simpler in structure than its flavour singlet analogue,
Eq.~\ref{Eq:ax_anom}.  In LQCD, the computation of the $\pi$-meson proceeds by
hitting the vacuum state $|0\rangle$ with the quark bilinear pseudoscalar
operator as given on the right hand side of \eq{Eq:pcac}. This  creates
both pions and excited $\pi$-like states.  Let us consider such an operator
carrying negative charge
\begin{equation}
\label{Eq:state}
 \bar{u}(x)\gamma_5 d(x) |0\rangle = |\pi ^-\rangle _x +  \mbox{ excited states} 
\end{equation}
with up and down quark operators, $u$ and $d$ respectively. The propagator
in the pseudoscalar meson channel from $x$ to $ x'$,
comprising the $\pi^-$-meson  and its excited friends,  would hence read like this:
\begin{equation}
\label{Eq:pi_prop}
 G^{(8)}_{\mbox{PS}}(x, x') = \langle 0| \bar{d}(x')\gamma_5 u(x')  \bar{u}(x)\gamma_5 d(x) |0\rangle  \;.
\end{equation}
This vacuum expectation value is most easily computed by Wick
contracting the two valence quark operators with matching flavour, and the result can be simply expressed
in terms of the two quark propagators, $P_u = M_u^{-1}(x,x')$ and
$P_d = M_d^{-1}(x,x')$ as follows:
\begin{equation}
\label{Eq:pion_prop}
 G^{(8)}_{\mbox{PS}}(x, x') = \mbox{Tr} [P^{*}_u(x,x')P_d(x,x')]\; ,
\end{equation}
where the trace is to be taken both in spin and colour space.
In LQCD, the quark propagators (from source $x_0$ to sink $x'$)
 can be computed in the background
field of the QCD ground state (the vacuum) by solving the
lattice Dirac equations
\begin{equation}
\label{Eq:dirac_eq}
 M_D(x')P(x,x') = \delta(x-x_0)\; ,
\end{equation}
where $M_D$ is the discretized form of the Dirac operator
including the gluon interaction and the {\it r.h.s.} denotes  a point
source located at lattice site  $x_0$,
see  \fig{Fig:g8}.
%FIG3
 This is done by applying modern iterative
solvers, like BiCGstab~\cite{bicgstab}. Note that
\begin{itemize}
\item $M_D\{U\}$ depends on the gluonic field configuration (i.e.  the QCD
  vacuum configuration $\{U\}$ that has been produced independently by a
  suitable sampling process such as Hybrid Monte Carlo~\cite{kennedy,sesam}) 
\item the efficiency of the solver is governed by the (fluctuating) condition
  number of the large sparse matrix $M_D\{U\}$~\cite{bicgstab},\cite{ssor}.
\end{itemize}
The simulation of the QCD vacuum on the lattice being a nonperturbative
computation, the gluonic lattice vacuum configurations comprise  the entirety of vacuum
polarization effects, i.e.  {\it any conceivable } fluctation from creation
and annihilation of quarks and antiquarks (sea quark effects). So in
\fig{Fig:g8} all vacuum polarization diagrams  as depicted in \fig{Fig:pola} are
understood to be included.
%FIG4
We note in passing, that the numerical computation of the complete inverse of
the lattice Dirac operator is prohibitively expensive: its computational
effort grows with lattice volume, which in todays simulations is anywhere
between $16^4$ and $64^4$ space-time lattice sites! So, in practical
calculations of the $\pi$-meson, one restricts the source, $x_0 $, to just a
few space points in a particular time slice, say in $t = 0$. The task for
calculation is then the pseudoscalar two-point function:
\begin{equation}
\label{Eq:8_corr}
 G^{(8)}_{\mbox{PS}}(x_0, x') = \Big < \mbox{Tr}
[P^{*}_u(x_0,x')P_d(x_0,x')]\Big >\; ,
\end{equation}
where the brackets denote the average over an ensemble of some hundred gauge
field vacuum states.  Note that it is enough in practice to compute but a few
columns of the entire inverse Dirac matrix per configuration for achieving
sufficient accuracy of the estimate, \eq{Eq:8_corr}. Once this pseudoscalar
two-point function has been calculated the pion contribution is obtained by
removing the excited states from the pseudoscalar channel. For this purpose,
one first extracts the zero momentum contributions by summing over all
(spatial) $\vec{x}'$ in time slice $t'$:
\begin{equation}
\label{Eq:two_point}
 C^{(8)}_{\mbox{PS}}(t=0,t') := \Big < \sum_{\vec{x}'} \mbox{Tr}
[P^{*}_u(x_0,x')P_d(x_0,x')]\Big >\; .
\end{equation}

This latter two-point function at large enough time separations $t'$ between
source and sink will be dominated by the ground-state of the channel, i.e. the
pion. Prior to that, it contains a superposition of ground and excited states:
 \begin{equation}
\label{Eq:pion_prop_as}
C^{(8)}_{\mbox{PS}}(0,t') =  \sum_{i} c_i \exp (-E_i\,  t')
\stackrel{t \rightarrow \infty}{\longrightarrow} c_0 \exp (-m_{\pi}
t') \; .
\end{equation}
It is obvious that the strict localization of the source at the very lattice
site $\vec{x}_0$ overly induces excited state contaminations; hence a suitable
smearing of the source around this site will deemphasize such pollutions and
will help to achieve precocious ground state dominance. This is welcome
because the two-point correlator decreases exponentially and is prone to 
suffer large
statistical fluctuations at large values of $t'$.

The numerical task in exploiting \eq{Eq:pion_prop_as} is then to optimize the
smearing such as to achieve an early onset of a plateau: the latter is defined
as the $t'$-range over which $G^{(8)}_{\mbox{PS}}(0,t')$ decays with a single
exponential. This is tantamount to requiring the 'local pseudoscalar mass',
$m_{\mbox{PS}}(t')$, to remain constant:
\begin{equation}
\label{Eq:pion_asym}
m_{\mbox{PS}}(t') := - \partial_{t'}ln [C^{(8)}_{\mbox{PS}}(0,t')] \stackrel {!}{=} \mbox{const.} = m_{\pi}\; .
\end{equation}
%For illustration of the numerical quality of such flattening out, we 
%show in \fig{Fig:localmass}  the flavour octet pseudoscalar correlator from the SESAM simulation of
%unquenched  QCD, at their lightest sea quark mass~\cite{sesam}.

Smearing of sources or sinks is achieved by a diffusive iteration~\cite{sesam}
process: think of replacing pointlike sources or sinks by
\begin{equation}
\label{Eq:diffusing}
\delta(x_0) = \phi^{(0)}(x_0) \longrightarrow \phi^{(N+1)}(x)\; ,
\end{equation}
where $\phi^{(N)}(x)$ is peaked around $x_0$  and is constructed  by the recursion
\begin{equation}
\label{eq:smearing}
\phi_s^{(i+1)}(x) = \frac{1}{1+6\alpha}\left[\phi_s^{(i)}(x,t) +\alpha\sum_{\mu}
 \phi_s^{(i)}(x + \mu )^{\mbox{{\tiny  p.t.}}}\right]\; .
\end{equation}
The index `p.t.' stands for `parallel transported', and the sum extends over
the six spatial neighbours of $x$.  As a result the entire series of smeared
'wave functions' $\phi^{(i)}, i = 1, 2, \cdots , N-1$ is  gauge
covariant under local gauge transformations.  Quark operators are computed
after $N=25$ such smearing steps, with the value $\alpha = 4.0$.  The smearing
procedure is applied to meson sources as well as to sinks for connected and
disconnected diagrams.  
%It should be kept in mind that when applying the
%stochastic estimator technique on the disconnected diagrams one should
%initialize the diffusive generation of sources by choosing a $Z_2$-noise
%nonlocal source, $\phi^{0}(x)$, in order to ensure  the correct normalization
%between connected and disconnected diagrams under smearing.

%\begin{figure}
%\begin{center}
%\leavevmode
%\mpicplace{9 cm}{3 cm}
%\end{center}
%\caption{
%Local mass from  the flavour octet  pseudoscalar
%correlator, $C^{(8)}_{\mbox{PS}}(t')$,
%at the lightest sea quark mass of the unquenched QCD SESAM
%simulation.}
%\label{Fig:localmass}
%\end{figure}

\subsection{Flavour singlet sector}
\label{Sec:singlet}
In the case of $SU(3)$ flavour symmetry the appropriately normalized flavour
singlet pseudoscalar operator reads
\begin{equation}
 O^{(0)}(x) =
\frac{1}{\sqrt{3}} \sum_{i} \bar{q}_i(x) \gamma_5 q_i(x)\; .
\end{equation}

\begin{figure}[ht]
\begin{center}
\includegraphics[width=.5\textwidth]{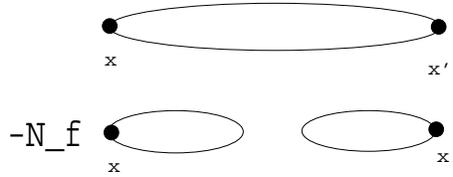}
\end{center}
\caption[]{The connected and OZI-rule violating ('disconnected')
  contributions to the $\eta'$ propagator $G_{\mbox{PS}}^{(0)}$ from
  \eq{Eq:eta_prop}. Note that these diagrams keep only track of the valence
  quark lines: all gluon exchanges due to the vacuum background field have
  been omitted for clarity of the graph.  It goes without saying that in
  unquenched QCD, the gluon propagators implicitly induce sea quark loops as
  in \fig{Fig:pola}.}
\label{Fig:eta'}
\end{figure}

Here the sum extends now equally over all (mass degenerate $u$, $d$,
and $s$) quark flavours.  The flavour singlet pseudoscalar meson
propagator is estimated from the average 
\begin{equation}
\label{Eq:eta_prop}
 G^{(0)}_{\mbox{PS}}(x, x') = \Big < O^{(0)\dagger}(x') O^{(0)}(x)
\Big >  \; .
\end{equation}
In this case the Wick contraction now induces both connected and
disconnected (two-loop) diagrams, $D$,
\begin{equation}
\label{Eq:2loop}
 D(x,x') = \Big < \mbox{tr} [O^{(0)\dagger}(x')] \mbox{tr} [ O^{(0)}(x)]
\Big >  \;,
\end{equation}
such that finally
\begin{equation}
\label{Eq:eta_propp}
 G^{(0)}_{\mbox{PS}}(x, x') = G^{(8)}_{\mbox{PS}}(x, x') - N_f D(x,x')  \;,
\end{equation}
as depicted in \fig{Fig:eta'}. Note that at this stage the trace operations in
\eq{Eq:2loop} refer to spin and colour space only and the Wick contraction
procedure leads to the weight $N_f = 3$ of the disconnected relative to the
connected contribution, the latter being identical to the expression given in
\eq{Eq:two_point}. The relative minus sign is due to Fermi statistics of the
quark operators.

We note in passing, that connected and disconnected
correlators come along with intrinsic positive sign such that the
above minus sign leads to a steepening of $C^{(0)}(t')$ w.r.t.
$C^{(8)}(t')$, iff $D$ shows a soft drop in $t'$.  
%It is obvious that  
%a smaller  slope in $D$ than in $C^{(8)}$ will increase $m_{\eta'}$ as
%the number of active sea quark  flavours  is raised.

Let us mention that the terminology ``two-loop correlator'' might be
misleading: it refers to the fact that we are dealing here with two (valence
quark) loop insertions of the $\gamma_5$ operator. The unquenched QCD vacuum
includes of course a host of sea quark loops that are {\it implicitely} taken
into account when sampling unquenchced QCD vacua in a LQCD simulation. They
would enter in form of vacuum polarization effects on the very gluon
propagators as depicted in \fig{Fig:pola}.

\begin{figure}[b]
\begin{center}
\includegraphics[width=.5\textwidth]{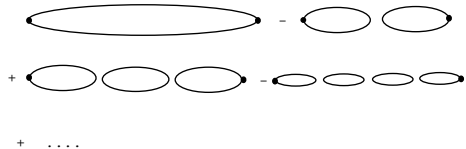}
\end{center}
\caption[]{Pictorial receipe for  setting up    the $\eta's$ propagator in quenched
  QCD as an explicit  multiloop  expansion.}
\label{Fig:geom_series}
\end{figure}

It is illuminating to consider the mechanism for the origin of the singlet
octet mass gap in a model to real QCD: let us start in the $N_c = \infty $
world, where the $U_A(1)$ symmetry holds and flavour octet and singlet mesons
are degenerate, $m_0 = m_8$.  This degeneracy is only broken by the higher
quark loop contributions which are suppressed by inverse powers in $N_c$.  It
is hence straightforward to approximate the $\eta'$ propagator, starting out
from this scenario, as an {\it explicit} multiloop expansion of quark
antiquark annihilation diagrams in the pure gluonic background field of
quenched QCD, as illustrated in \fig{Fig:geom_series}.  The actual calculation
can be greatly simplified by introducing an effective loop-loop coupling, $\mu
_0^2$. In momentum space the $\eta'$-propagator then takes the form of a
geometric series:
\begin{equation}
\label{Eq:geom_series}
P_{\eta '} = \frac{1}{p^2 + m_8^2}\Bigg ( 1 + \sum_{l=1}^{\infty} \bigg [\frac{- \mu _0^2}{p^2+m_8^2}\bigg ] ^l \; \Bigg)\; .
\end{equation}
The series is readily
summed up with the result:
\begin{equation}
\label{Eq:quenched_eta}
P_{\eta'} =  \frac{1}{p^2 + (m_8^2 + \mu_0^2)}\; ,
\end{equation}
which reveals that the effective loop-loop  coupling is nothing but the mass gap
between singlet and octet pseudoscalar masses:
\begin{equation}
\label{Eq:gap}
\Delta m^2 =: m_0^2 - m_8^2 =  \mu_0^2  ,
\end{equation}

We emphasize that in this effective model it is only by {\it summation of the
  entire series to all orders in $l$} that we arrive at a simple pole
expression for the singlet propagator in momentum space and hence at a simple
exponential falloff in time of the two-point correlator.  Note that this
summation happens automatically in an unquenched QCD situation where all
multiquarkloop effects are already implicitly included in the very gauge field
vacuum configuration!

\section{The  real world with   $n\bar{n} $ $s\bar{s}$ mixing }
\label{Sec:realistic}
So far we have been leading a rather academic discussion in an SU(3) symmetric
world.  We would like next to describe the $\eta$/$\eta '$ problem in a more
realistic setting. This opens up a pandora box of four parameters for decay
matrix elements plus the mixing angle among isosinglet particle states.  Two
degrees of freedom in the decay constants can be characterized as mixing
angles~\cite{leutwyler}.

\subsection{Phenomenological approach: alignment hypothesis}
\label{alignment}
In order to ease the interpretation of flavour breaking effects in the
isosinglet sector, Feldmann et al~\cite{fks,kroll} have proposed some time ago
an intuitive scheme to describe the $\eta - \eta'$ phenomenology by use of a
quark state Fock state representation in conjunction with an alignment
assumption.  The latter reduces the number of parameters to {\it two}
couplings plus {\it a single } mixing angle of the particle states in that
Fock space.  Let us briefly review their scenario: they proceed from the two
isosinglet states
\begin{eqnarray}
\label{Eq:fock}
|\eta_n\rangle &=&\Psi_n |u\bar{u} + d\bar{d}\rangle/\sqrt{2} + \dots\\
|\eta_s\rangle &=&\Psi_s |s\bar{s}\rangle + \dots \quad ,
\end{eqnarray}
where  $\psi_{n,s} $ denote  light cone wave functions~\footnote{
There dots in \eq{Eq:fock} represent higher Fock states, including $|gg\rangle
$ and components from the heavy quark sectors, such as $\bar{c}c\rangle$.}.
The key assumption of their phenomenological approach   is their {\it hypothesis  of
alignment} of the current matrix elements:
\begin{equation}
\langle 0|\bar{q}_i\gamma_5 \gamma_{\mu} q_i |\eta_j\rangle = \delta_{ij}f^i
p_{\mu} \; , i,j = n,s \; .
\label{Eq:alignment}
\end{equation}

Due to the QCD interactions  the physical $\eta$ and $\eta'$ states 
will turn out to  be mixtures of these Fock states:
\begin{eqnarray}
\left( \begin{array}{c}
|\eta\rangle\\
|\eta'\rangle
\end{array} 
\right)=
U(\phi)
\left( \begin{array}{c}
|\eta_n\rangle\\
|\eta_s\rangle
\end{array} 
\right):=
\left( 
\begin{array}{cc}
\cos\phi   & -\sin \phi \\
\sin \phi  & \cos \phi
\end{array}
\right)
\left( \begin{array}{c}
|\eta_n\rangle\\
|\eta_s\rangle
\end{array}
\right)
 \; .
\label{Eq:mixture}
\end{eqnarray}

As a consequence of the  alignment hypothesis,  \eq{Eq:alignment},  the four decay constants, $f^i_P$
\begin{equation}
\label{Eq:decays}
\langle 0| j_{\mu 5}^i|P \rangle= \langle 0|\bar{q}_i\gamma_5 \gamma_{\mu}q_i|P\rangle \equiv f^i_Pp_{\mu} \quad 
i = n, s; \quad    P = \eta , \eta' 
\end{equation}
can  be expressed in  terms of $f^n, f^s$ and the very   rotation angle
$\phi$~\cite{fks}:

\begin{eqnarray}
\left( 
\begin{array}{cc}
f^n_{\eta}   & f^s_{\eta} \\
f^n_{\eta'} & f^s_{\eta'}
\end{array}
\right) =
U(\phi)
\left( 
\begin{array}{cc}
f^n   & 0 \\
0  & f^s
\end{array}
\right)
 \; .
\label{Eq:f_mixture}
\end{eqnarray}

By feeding  the anomalous PCAC  relation 
\begin{equation}
\partial^{\mu} \bar{q}_i\gamma_5\gamma_{\mu}q_i = 2 m_i \bar{q}_i \gamma_5 q_i
+ \tilde{Q}
\label{Eq:an_pcac}
\end{equation}
into the four   decay matrix elements
\begin{equation}
\label{Eq:fock_decays}
\langle 0|\partial ^{\mu} j_{\mu 5}^i|\eta \rangle = f^i_{\eta} m_{\eta}^2,
\quad \langle 0|\partial ^{\mu} j_{\mu 5}^i|\eta' \rangle = f^i_{\eta'}
m_{\eta'}^2;  \quad (i = n, s) \; , 
\end{equation}
the authors of ref.\cite{fks,kroll,feldmann} arrive at the mass matrix mixing
relation
\begin{eqnarray}
\left( 
\begin{array}{cc}
\mu_{nn}^2 + \frac{\sqrt{2}}{f^n}\langle 0|\tilde{ Q}|\eta _n\rangle   & \frac{1}{f^s}\langle 0|\tilde{ Q}|\eta _n\rangle \\
\frac{\sqrt{2}}{f^n}\langle 0|\tilde{Q}|\eta _s\rangle & \mu_{ss^2}+ \frac{1}{f^s}\langle 0|\tilde{ Q}|\eta _s\rangle \\
\end{array}
\right) =
U^{\dagger}(\phi)
\left( 
\begin{array}{cc}
m^2_{\eta}  & 0 \\
0  & m^2_{\eta'}
\end{array}
\right)
U(\phi)
 \; ,
\label{Eq:mass_matrix}
\end{eqnarray}
with a single mixing angle that is in accord with Fock state mixing.
In  \eq{Eq:mass_matrix} the abbreviations $\mu ^2$ stand for the explicit chiral symmetry
breaking terms from nonvanishing quark masses, $m_i$:
\begin{equation}
\mu_{ii}^2 := \frac{2m_i}{f_n}\langle 0|\bar{q_i}\gamma_5 q_i |\eta_i \rangle;
\quad i = n,s \; .
\label{Eq:chiral_mass}
\end{equation}
For consistency of the approach, they have to postulate the symmetry of the
mass matrix, which sets a constraint among the matrix elements $\langle
0|\tilde{ Q}|\eta _i\rangle; \; i = n,s \; .$ 
 
The l.h.s. of \eq{Eq:mass_matrix} being a continuum relation, great care must
be exercised in renormalizing the operators. Since the underlying anomalous
PCAC relation, \eq{Eq:an_pcac} is actually to be seen as a Ward-Takahashi
identity from chiral rotations, the standard Wilson
lattice fermions provide a problematical scheme for regularizing
the matrix entries in \eq{Eq:mass_matrix}, with the Wilson lattice fermions
lacking chiral symmetry at finite lattice spacing~\cite{bochiccio}. This
complex situation would be largely improved when dealing with Ginsparg-Wilson
fermions in unquenched QCD, whose lattice chiral symmetry prevents undesired
divergencies from operator mixings~\cite{lattice_WV}.  On the lattice, the
mass$^2$ entries to the matrix \eq{Eq:mass_matrix} are best accessible from
propagator studies in the singlet channels~\cite{kilcup,ukqcd,michael,michael_rev}.

We would like to explain next, that the idea of Fock state mixing of
eigenstates can in principle be pursued very naturally from a variational
approach on the lattice, once we have fully unquenched QCD vacuum configurations in the
future, without making reference to an alignment hypothesis.

\subsection{ Mixing singlet mesons on the lattice} 
\label{lattice_mixing}
On the lattice, we can prepare the the Fock states $|\eta_n\rangle$ and $
|\eta_s\rangle$ by hitting the vacuum at some (smeared) source location, say
$x=(0,0,0,0)$, with the bilinear isosinglet operators,
\begin{equation}
{\cal O}_i(0)|0\rangle  =  \bar{q}_i(0)\gamma_5 q_i(0)|0\rangle , \quad i=n,s
\label{Eq:bilinear}
\end{equation}
and by waiting long enough to see the ground states emerge. 

In the sense of a variational method, one can next work with a superposition
of these operators to achieve earlier exponential decay of the correlator
and/or lowering of the ground state, the $\eta$-meson:
\begin{equation}
{\cal O}_{\alpha } = \cos \alpha\; {\cal O}_n - \sin \alpha\;
{\cal O}_s \; .
\label{Eq:var_ansatz}
\end{equation}
Suppose this variational ansatz delivers an immediate plateau for some value
of the variational parameter, $\alpha = \alpha^* $: in that case we have
actually hit an eigenmode  of the interacting system and the Fock state
mixing angle is $\phi = \alpha^*$!  Since one expects little mixing with
higher Fock states, one could then recover the $\eta'$ ground state from the
orthogonal combination
\begin{equation}
{\cal O}_{\alpha }^{\perp} = \sin \alpha\; {\cal O}_n + \cos \alpha\;
{\cal O}_s \; .
\label{Eq:var_ansatz_per}
\end{equation}

In  general the  state mixing information is encoded  more deeply in the correlation
matrix (still in operator quark content basis)
\begin{eqnarray}
{ \cal C }(t) = 
\left( \begin{array}{cc}
C_{nn}(t) - 2D_{nn}(t) & \sqrt{2}D_{ns}(t)\\
\sqrt{2}D_{sn}(t)           & C_{ss}(t) - D_{ss}(t)
\end{array}
\right)
 \; . 
\label{Eq:corr_mat}
\end{eqnarray}
Note that the weight factors $2$ and $\sqrt{2}$ in this relation originate
from the combinatorics when Wick contracting the degenerate $u$ (with $\bar{u}$)
and $d$ (with $\bar{d}$) quark operators contained ${\cal O}_n$
\begin{equation}
\bar{q}_i\gamma_5 q_n := \frac{1}{\sqrt{2}}(\bar{u}\gamma_5 u +
\bar{d}\gamma_5 d ) \; . 
\label{Eq:def_singl}
\end{equation}

If plateau formation does not occur immediatley, at $t_p = 0$, one needs to
follow the time evolution up to a later time step when the higher excitations
have died out, and ascertain the plateau condition within the time interval
$t_p \rightarrow t_p+\Delta t$ in form of a generalized eigenvalue
problem~\cite{wolff}:

 \begin{eqnarray}
{\cal C}(t_p+\Delta t)
\left( \begin{array}{c}
+\cos \alpha \\
-\sin \alpha 
\end{array} 
\right)=
\exp (-m_{\eta}\Delta t) \;  \; {\cal C}(t_p)
\left( 
\begin{array}{c}
+\cos \alpha \\
-\sin \alpha 
\end{array}
\right)
 \; .
\label{Eq:gen_eigen_eta}
\end{eqnarray}
The  problem can be recast into a symmetric form
\begin{equation}
{\cal C}^{-1/2}(t_p){\cal C}(t_p+\Delta t){\cal C}^{-1/2}(t_p)\cdot \vec{\eta_{\alpha}}=
\exp (-m_{\eta}\Delta t) \cdot 
\vec{\eta_{\alpha}}
 \; 
\label{Eq:gen_eigen_sym}
\end{equation}
with the definition
\begin{eqnarray}
\vec{\eta_{\alpha}}(t_p)=
{\cal C}^{+1/2}(t_p)
\left( \begin{array}{c}
+\cos \alpha \\
-\sin \alpha 
\end{array} 
\right)
\end{eqnarray}

According to the symmetric plateau condition, \eq{Eq:gen_eigen_sym}, the
direction of the ground state vector, $\vec{\eta_{\alpha}}(t_p)$, remains
fixed under further time displacements $\Delta t$ through the transfer
operator, ${\cal T}(\Delta t)= {\cal C}^{-1/2}(t_p){\cal C}(t_p+\Delta t){\cal
  C}^{-1/2}(t_p)$. Hence the $\eta$ mixing angle in the quark state basis is
given by the direction of eigenvector $\vec{\eta_{\alpha}}(t_p)$ at the onset of the
plateau  and might differ appreciably from the variational angle $\alpha$.
The problem \eq{Eq:gen_eigen_sym} being symmetric, the $\eta'$ is retrieved as
the second, perpendicular eigenstate of ${\cal T}(\Delta t)$.

It might be illuminating to put \eq{Eq:gen_eigen_sym} into perspective with
the previous plateau condition, \eq{Eq:pion_asym}: by Taylor expanding of
${\cal C}(t_p+\Delta t)$ to lowest order in $\Delta t$ one can readily recast
\eq{Eq:gen_eigen_sym} into a local condition, with a 'logarithmic derivative'
matrix construct, ${\cal C}^{-1/2}(t_p){\cal C}'(t_p){\cal C}^{-1/2}(t_p)$:
\begin{equation}
{ \cal C }^{-1/2}(t_p){ \cal C }'(t_p){ \cal C }^{-1/2}(t_p)\cdot  \vec{\eta_{\alpha}}(t_p)=
-m_{\eta}\cdot \vec{\eta_{\alpha}}(t_p) \; .
\label{Eq:gen_plat}
\end{equation}

In principle, the phenomenological alignment postulate of \eq{Eq:alignment}
can be tested on the lattice by computing another, nonsymmetric set of
correlators
\begin{equation}
\tilde{{\cal C}}_{ij}(t) =
\langle 0|\partial
^{\mu}\bar{q}_i(t) \gamma _5 \gamma _{\mu} q_i(t)  \bar{q}_j(0) \gamma _5  q_j(0)
|0\rangle \; ,
\end{equation}
which allows to extract the various decay matrix elements.  Let us emphasize
in concluding this section that the scenario presented here relates to fully
unquenched three flavour QCD; according to the discussion in \ref{Sec:singlet}
the mass plateau conditions can only be anticipated if the quark contents of
the bilinear operators in \eq{Eq:bilinear} are equally present  in the quark
sea~\footnote{see the discussion on the partially quenched approach in
section \ref{Sec:partially_quenched}}.

\section{The three computational bottlenecks}
\label{signal}
In this section we shall focus  in more detail on the numerical obstacles  that
one encounters when carrying out the above program. These difficulties  are
computationally much more severe than in nonsinglet spectroscopy and hence
require the development of advanced algorithms and numerical techniques:

{\bf 1. Present limitations in sea quark flavours.}  It was only in the second
half of the nineties that LQCD had developed the means to tackle first
semi-realistic large scale simulations of QCD beyond the valence (or
`quenched') approximation.  The reason is that Wilson-like discretizations of
the Dirac operator are mandatory once we wish to deal with strong interaction
situations sensitive to flavour symmetry. But such Wilson-like lattice
fermions need extremely large supercomputer resources when it comes to
actually simulate the effects of dynamical sea quarks, i.e. to sample
unquenched CCD vacuum configurations.  The hybrid Monte Carlo algorithm (HMC)
is a nonlocal sampling technique which was developed and improved ever since
the late eighties~\cite{kennedy}; it continues to be the workhorse in all of
the major QCD simulation projects with Wilson-like
fermions~\cite{sesam,cppacs,ukqcd,qcdsf}.  For technical
reasons, however, HMC has the shortcoming of being limited to even numbers of
sea quark flavours.  As a consequence, all the above projects neglected
dynamical $s$-quarks in working with mass degenerate $u$ and $d$ sea quarks
only, i.e.  considered an $SU(2)$ symmetric sea of quarks only.  It is to be
hoped that the next generation of large-scale QCD simulation with
Teracomputers will overcome this limitation with new sampling
algorithms.

{\bf 2. Trace computations.}  We have seen that the peculiarity of flavour
singlet objects lies in the annihilation of the valence quark lines into the
flavour blind gluonic soup which implies the need to compute disconnected
diagrams.  But inspection of \eq{Eq:2loop} shows that the
$\vec{x'}$-summation from  momentum zero projection requires the evaluation
of the  trace of the inverse Dirac operator.  This task is
definitely beyond the reach of modern linear equation solvers for matrices of
rank $10^6$ and more.  Traditionally  stochastic estimator techniques 
(SET) have been the popular  workaround~\cite{stoch,struckmann}.

\paragraph{\bf 3. Noisy signals.}
The singlet spectrum being inherently determined by the physics of the quantum
fluctuations of the QCD vacuum, the two-loop correlators are bound to suffer
from a serious noise level.  This calls for the use of noise reduction methods,
in order to circumvent the need for  overly {\it large ensembles of vacuum field
  configurations}. With this motivation low mode expansion methods have been
proposed recently with considerable success to replace stochastic estimator
techniques\cite{kilcup,tea}.

\subsection{Stochastic estimate of fermion loops}
\label{Sec:stochastic}
The basic idea of stochastic estimation of the momentunm zero projected loop
operator in \eq{Eq:2loop}, $\sum_{\vec x}\mbox{tr}\big( \gamma_5M^{-1}(\vec x,
t)\big) $, is very simple: instead of attempting to solve \eq{Eq:dirac_eq} $V$
on delta-like sources, one introduces socalled stochastic volume
sources which are completely delocalized, on the entire lattice volume V:
\begin{equation}
\xi = (\xi_1, \xi_2, \dots , \xi_V)\; .
\label{Eq:rand_source}
\end{equation}
The components $x_i$ are chosen to be real random numbers sampled from a
normal distribution $N(x_i) = 1/\sqrt{2\pi}\exp (-x_i^2/2)$ or socalled $Z_2$
noise which is nothing but equally distributed random numbers $\pm 1$.  One
solves the Dirac lattice equation on an ensemble of $N_s$ such random source
vectors, $\{ \xi^{\alpha} \}$, for each individual gauge configuration:
\begin{equation}
\zeta^{\alpha} =  \gamma_5 M_D^{-1}\xi^{\alpha} \; .
\label{Eq:dirac_stoch}
\end{equation}
and computes the ensemble average of
the  scalar products
\begin{equation}
\Big < (\xi, \zeta)\Big > _s \; := \;
\frac{1}{N_sV}\sum_{\alpha=1}^{N_s}\sum_{i=1}^V \xi^{*\alpha}_i
\zeta^{\alpha}_i \simeq \mbox{TR} \Big [\gamma_5
M_D^{-1}\Big ]      \; .
\label{Eq:av_stoch}
\end{equation}
One might phrase this procedure as an all-at-one-stroke method, at the expense
of nonclosed loops sneaking in, in form of contributions from $\Big [\gamma_5
M_D^{-1}\Big ]_{ij}$
with $i\neq j$.  These undesired  contributions are suppressed, however, due to
the 'componentwise orthonormality' relation
\begin{equation}
\Big < \xi_i^*\xi_j\Big >_s =    \frac{1}{N_s}\sum_{\alpha=1}^{N_s} \xi^{*\alpha}_i \xi^{\alpha}_j \stackrel{N_s\rightarrow \infty}{\longrightarrow}\delta_{ij}\; .
\label{Eq:orth}
\end{equation}
Note that \eq{Eq:av_stoch} readily allows for restricting the grand
trace operation over the entire space-time lattice, TR,
to any particular time slice, simply by truncating the $i$-summation to the
appropriate three-dimensional subvolume, such as to yield the estimator for
the momentum zero expression on time slice $t$, $\sum_{\vec x}\mbox{tr}\big(
\gamma_5M^{-1}(\vec x,t)\big) $.

In principle, these considerations hold for all kinds of insertions, not just
$\gamma_5$ like in our example. In any case the 'sneakers' are suppressed as
${\cal O}(N_s^{-1})$ terms, but with a strength that depends on the insertion.
The practical advantage of this stochastic procedure is, that in the instance
of $\gamma_5$ insertions one needs to compute only some few hundred solutions
to \eq{Eq:dirac_stoch} which is much less than ${\cal O}(V)$! But we should
keep in mind that this is achieved only at the expense of injecting additional
noise into the problem\footnote{We would like to point out that some
  authors~\cite{ukawa_elitz} have refrained from using noisy sources
  altogether by using a single 'solid' wall source with $\xi_i=1$ and relying
  on gauge symmetry to suppress the unwanted nonclosed loops, $\Big
  [\gamma_5 M^{-1}_D\Big ]_{ij}\;, \mbox{with} \, i \neq j$.  Actually this argument relies
  on Elitzur's theorem according to which the vacuum expectation value
  $\langle O\rangle$ vanishes for any such non gauge invariant operator $O$!
  However, if the sample of vacuum gauge fields is limited to some few hundred
  independent configurations only this procedure offers too little control
  over these systematic errors; nevertheless one could refine this variant by
  applying the 'solid' wall source idea on an ensemble of stochastic gauge
  copies of each individual vacuum configuration. To our knowledge this has
  never been attempted.}.

\subsection{Low eigenmode approximation to fermion loops}
\label{Sec:TEA}
We start with the eigenvalue problem for the socalled Hermitian
Dirac operator, 
\begin{equation}
Q_5:=\gamma_5 M_D \;,
\label{Eq:hermi}
\end{equation}
 which reads
\begin{equation}
Q_5\psi_i = \lambda_i\psi_i\; .
\label{Eq:eigen_val}
\end{equation}
The quark loop with $\gamma_5$ insertion at time slice $t$ is
simply expressed in terms of this eigensystem:
\begin{equation}
\label{Eq:eigen_exp}
Q_5^{-1}(t) =  \sum_i \frac{1}{\lambda_i}
\frac{ \langle \psi_i(t) | \psi_i(t) \rangle}
{ \langle \psi_i | \psi_i \rangle} \; .
\end{equation}
The disconnected two-loop correlator from \fig{Fig:eta'} then has the form
\begin{equation}
\label{Eq:eigen_twoloop}
D(t) = 
\sum_{t_0} \sum_i \frac{1}{\lambda_i}
\frac{ \langle \psi_i(t_0) | \psi_i(t_0) \rangle}
{ \langle \psi_i | \psi_i \rangle} 
\sum_j \frac{1}{\lambda_j}
\frac{\langle \psi_j(t_0+  t) | \psi_j(t_0+ t) \rangle}
{\langle \psi_j | \psi_j \rangle}\; , 
\end{equation}
where $t$ represents the time separation between source and sink and we have
exploited translational invariance by summing over all time locations of the
source, $t_0$. This summation is for free, once the eigenfunctions are
available\footnote{We mention that there might be other opportunities for
  spectral methods in LQCD with their potential to deal with all-to-all
  propagators: like the long standing string breaking problem or heavy-light
  meson scattering\cite{alltoall}.}.  The determination of the entire spectrum
being of course prohibitively expensive, we proceed by arranging the spectrum
in ascending order of $|\lambda_i|$ attempting to saturate the eigenmode
expansion, \eq{Eq:eigen_twoloop}, with its lowest modes. The intuitive
reasoning behind such a strategy is, that the underlying physics is expected
to be encoded in the infrared rather than ultraviolet modes.  After all, it is
well known that for Wilson fermions 15/16  of the spectrum is related
to the unphysical doublers that become frozen to order ${\cal O }(a^{-1})$ as
the lattice spacing $a$ is going to zero.
\begin{figure}[ht]
\begin{center}
\includegraphics[width=.7\textwidth]{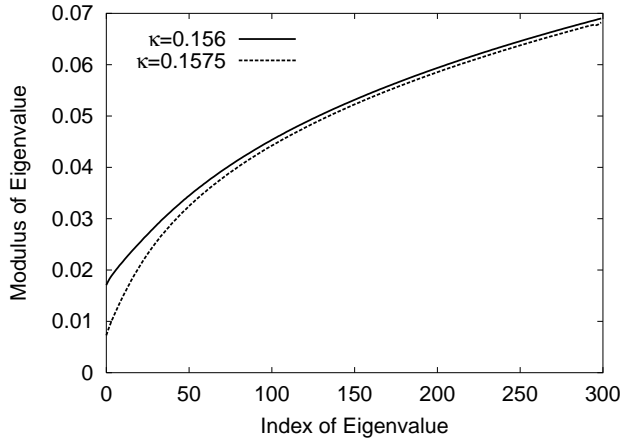}
\end{center}
\caption{
  Cumulative spectral distribution of the lowest eigenmodes at the smallest
  (lower curve) and largest (upper curve) sea quark masses of the SESAM
  project~\cite{tea}.}
\label{Fig:TEA_spectrum}
\end{figure}

\fig{Fig:TEA_spectrum} offers an impression on the cumulative distribution of
the lowest modes of the Dirac operator in the situation of the SESAM
simulation, at their smallest and largest sea quark masses. Note that the
masses of valence quarks (inserted into the Dirac operator) and sea quarks
(inserted into the hybrid Monte Carlo sampling of gauge field configurations)
are chosen to coincide; \fig{Fig:TEA_spectrum} therefore reflects the feedback
mechanism from dynamical effects onto the spectral representation as quark
masses are diminished.  \fig{Fig:TEA_spectrum} suggests that  the
important spectral activity is  mostly within   the lowest 50 modes.

But how many modes do suffice for saturating  \eq{Eq:eigen_twoloop}?
\begin{figure}
\begin{center}
\includegraphics[width=.7\textwidth]{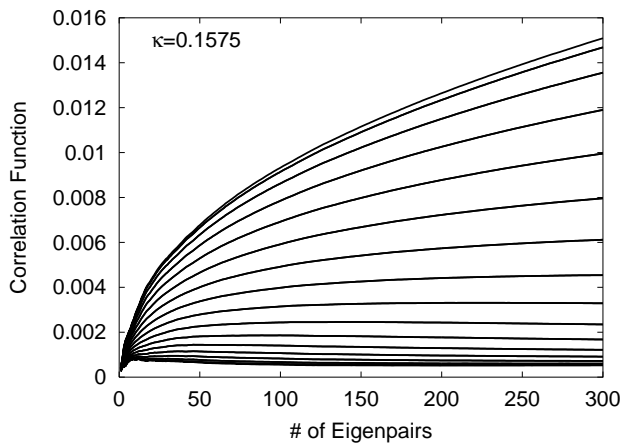}
\end{center}
\caption{
  Pion correlator in spectral approximation, plotted {\it versus} spectral
  cutoff, $\lambda _c$. Shown is a sequence of trajectories, as obtained on
  fixed values of $t$: the curves represent $C^{(8)}(t=1)$, $C^{(8)}(t=2)$,
  $C^{(8)}(t=3)$ {\it etc} , from top to bottom~\cite{tea}.  In this
  representation the justification of truncation is signaled by a flat
  $\lambda _c$-dependency.}
\label{Fig:TEA_goodness_pi}
\end{figure}

For the saturation of the two point correlators, one would anticipate the need
for {\it larger} values for the cutoff, $\lambda^c$, as one {\it diminishes} the time
separations towards a few lattice spacings.  This expectation is indeed
confirmed by the numerical results for the pion corrrelator as displayed in
\fig{Fig:TEA_goodness_pi}: The diagram shows a sequence of curves which -- in
descendig order -- refer to $C(t=1), C(t=2), \cdots , C(t=16)$, which are
plotted {\it versus} the spectral cutoff. For small values of $t$ the curves
keep rising with $\lambda _c$, while asymptotic saturation is found to be
better at larger $t$-values.

\begin{figure}
\begin{center}
\includegraphics[width=.7\textwidth]{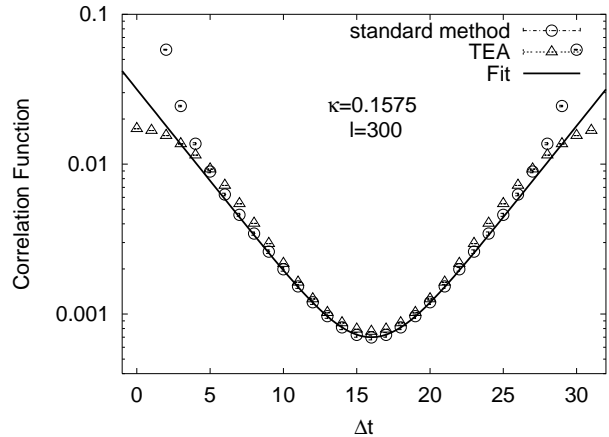}
\end{center}
\caption{
  Quality of the truncated eigenmode approximation at the smalles SESAM sea
  quark mass for the pion correlator, with a spectral cutoff, $\lambda _c=
  300$. Satisfactory agreement with the standard result from using iterative
  solvers is only reached at $t \simeq 10$.}
\label{Fig:TEA_pi_300}
\end{figure}

\begin{figure}[b]
\begin{center}
\includegraphics[width=.7\textwidth]{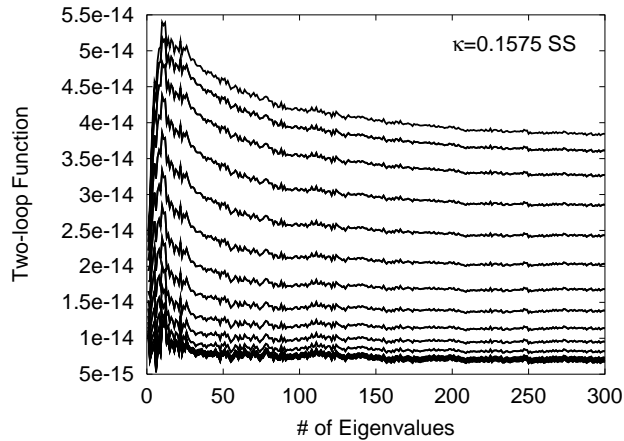}
\end{center}
\caption{
Quality of low eigenmode approximation as function of time separation for
the OZI rule violating term, $D(t)$. The curves represent (from top to bottom)
$C(1), C(2), \cdots , C(16)$~\cite{tea}.  }
\label{Fig:TEA_goodness_tl}
\end{figure}

In \fig{Fig:TEA_pi_300}, we plotted the pion correlator from a truncated
eigenmode approximation (TEA) with cutoff $\lambda_c = 300$ and from standard
iterative solvers. There remain noticeable differences in the two curves for
this value of the cutoff.  Therefore, one would refrain from using TEA for
connected hadron propagators unless really needed.  The situation {\it
  w.r.t.}  $\lambda^c$ turns out to be much more favourable with the
disconnected piece, $D(t)$, as shown in \fig{Fig:TEA_goodness_tl}.

\begin{figure}
\begin{center}
\includegraphics[width=.7\textwidth]{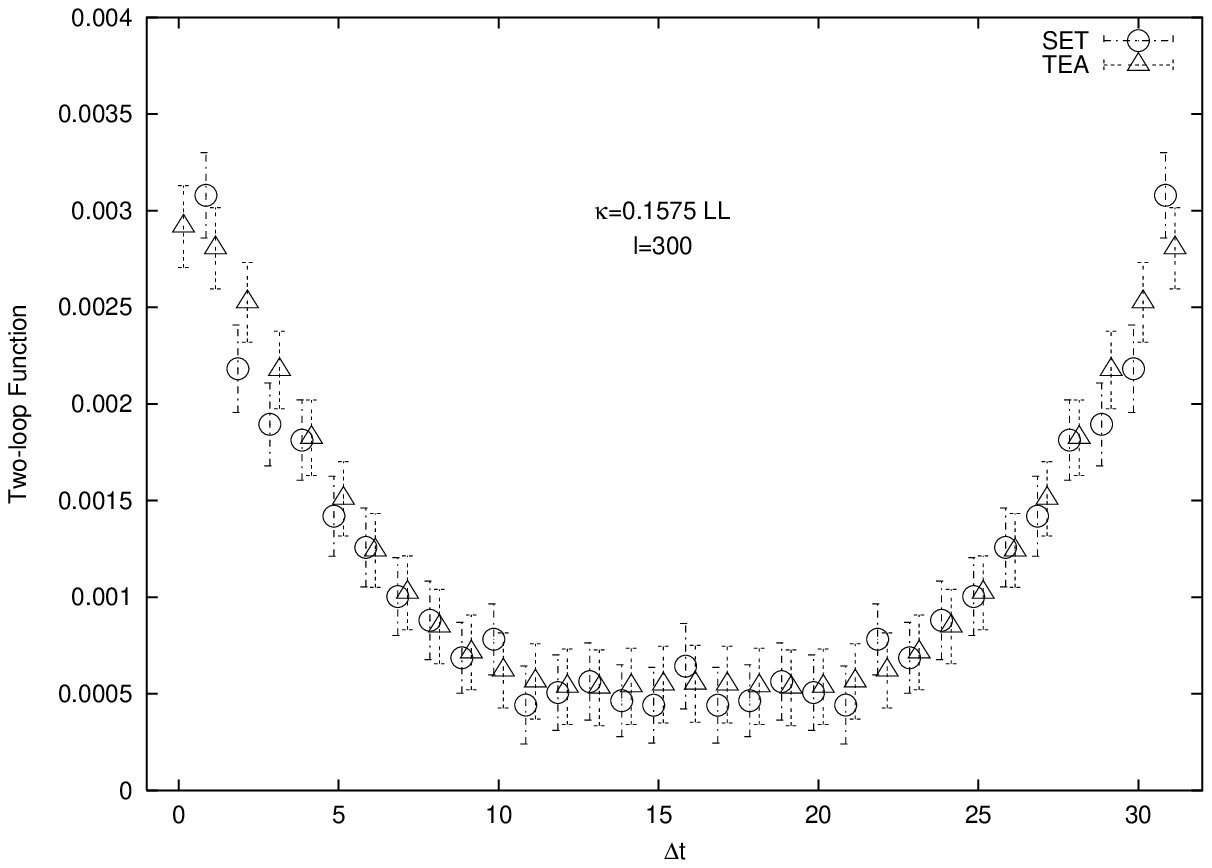}
\end{center}
\caption{
  The two-loop correlator, $D(\Delta t)$, as obtained from SET~\cite{struckmann} (circles) and TEA
  (triangles) in comparison. Local sources and sinks have been used and $N_s=
  300$ and $\lambda _c = 300 $ were chosen~\cite{tea}. }
\label{Fig:comparison}
\end{figure}

How well the idea of saturation actually works out quantitatively is
illustrated in \fig{Fig:comparison}, where we compare the two-loop correlator
in the truncated eigenmode approximation (TEA) against the results from a
previous comprehensive study with the estimation by an ensemble of stochastic
sources~\cite{struckmann}.  We find that the contributions from the lowest 300
eigenmodes agree remarkably well with the SET computations which are based
here on 400 source vectors, on a $16^§\times 32$ lattice in the sea quark mass
range covered used by the SESAM project! The net computational effort behind
the two sets of data points in \fig{Fig:comparison} is about equal, yet the
TEA data among themselvesa appear to be less fluctuating.

This positive message from \fig{Fig:comparison} bears substantial promise for
future lattice studies with lighter quark masses, as SET is doomed to degrade
when penetrating more deeply into the more chiral regime while the {\it
  truncated eigenmode approximation} (TEA) will thrive: (a) the lowest modes
will dominate all the more as $|\lambda_{min}|$ approaches zero; (b) modern
Arnoldi eigensolvers have been made very efficient in  present day LQCD
simulations~\cite{neff_solver} and will not deteriorate for lighter quarks as
long as one sticks to fixed lattice volumes.  Conversely, iterative solvers on
stochastic sources will suffer substantial losses in convergence rate, once
the condition number of the Dirac matrix, $c :=|\lambda_{max}/\lambda_{min}|$
starts exploding.

\subsection{Navigating between Scylla and Charybdis} 
\label{scylla}
There is another lesson from \fig{Fig:comparison}: we find the noise level in
$D(t)$ to be rather independent of $t$, i.e.  of signal size. This is of
course in line with the fact that we are hunting for vacuum fluctuations
proper. It therefore appears impracticable to achieve ground state projection
in the very standard fashion by increasing $t$: the $\eta'$ correlator being
the imbalance between connected and two-loop signals one is to keep track of a
signal in form of a {\it rapidly diminishing} difference between two
quantities, with one being at {\it fixed noise} level. Given this situation we
are challenged by complying to two conflicting needs: on one hand large $t$
values are needed to deflate excited state contaminations and on the other
hand the struggle for an acceptable signal-to-noise ratio impedes working in
the large-$t$ regime.  Our strategy to deal with this conundrum of the $\eta'$
correlator, \eq{Eq:eta_propp} is to contrive a procedure for circumventing
this conflicting requirements. The receipe is the following:
\begin{itemize}
\item suppress the  excited states from  $C_{PS}^{(8)}(\Delta t)$ by using an
  exponential fit from  the window of  its  mass plateau;
\item deemphasize  the short range noise level  inside  $D$ by means of the
  truncated eigenmode expansion, \eq{Eq:eigen_twoloop}.
\end{itemize}
This method could also be characterized through an intermediate 'synthetic
data stage' since the actual ground state analysis for the singlet correlator
is organized as a sequence of three steps:
\paragraph{\bf (a) extract}
 an exponential fit to the lattice data
for $E(\Delta t) = C_{PS}^{(8)}(\Delta t)$,
\paragraph{\bf (b) combine}
this {\it fit curve}, $E(\Delta t)$, with the {\it lattice data} for $D(\Delta t)$ into
the 'synthetic' data set for $C_{PS}^{(0)}(\Delta t):= E(\Delta t) -
N_fD(\Delta t)$, in the spirit of \eq{Eq:eta_propp},
\paragraph{\bf (c) establish } a mass plateau in $C_{PS}^{(0)}(\Delta t)$ and perform
another exponential fit to $C_{PS}^{(0)}(\Delta t)$ within the plateau region
in
order to determine the singlet pseudoscalar mass.\\

\begin{figure}[ht]
\begin{center}
\includegraphics[width=.7\textwidth]{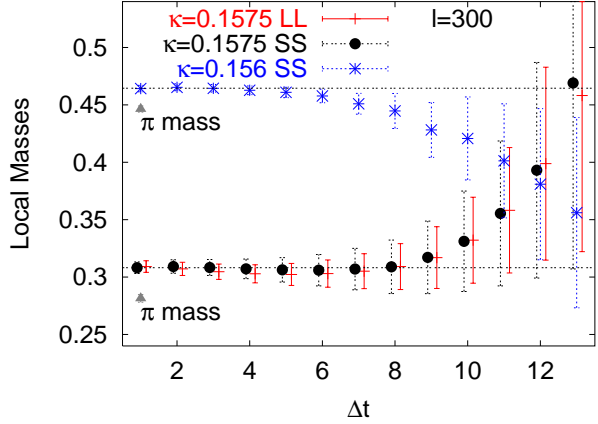}
\end{center}
\caption{
  Mass plateau formation in the $\eta'$ (singlet) correlator after dual
  filtering with TEA, for two different values of hopping parameters (sea
  quark masses) versus time separation between source and sink, $\Delta
  t$~\cite{tea}.  The plot shows the respective lattice values for the pion
  (nonsinglet) mass for comparison.  The effect of smearing onto the signal is
  also illustrated: SS and LL refer 'to smeared source and sink' and 'local
  source and sink', respectively.  }
\label{Fig:dual_filtering}
\end{figure}

We find this {\it dual filtering} approach to operate very well in the
two-flavour case, on the SESAM QCD configurations.  This is illustrated in
\fig{Fig:dual_filtering} which manifests very clearly a plateau in the local
mass plot of the $\eta'$-correlator. The onset of plateau formation is
precocious in the sense that we still maintain very precise signals. Therefore
the statistical accuracy is sufficiently high to resolve the mass gap between
the singlet and nonsinglet masses, whose lattice values for these sea quark
masses are also included in the plot, being indicated with the symbol '$\pi$'.

Let us emphasize that we consider the upshot of precocious asymptotia in the
$\eta'$-correlator as an essential outcome of our analysis; it corroborates
that the nonperturbative lattice simulations take effect in implicitly
creating the mass gap between singlet and nonsinglet states, much in line with
the perturbative reasoning from \eq{Eq:geom_series}: indeed, the simulation
data for the two-loop correlator, $D(t)$, turn out to be the {\it difference
  between just two exponentials} rather than a multiexponential superposition.
This will enable us to perform a rather detailed physics
analysis in the singlet pseudoscalar channel.

\begin{figure}[t]
\begin{center}
\includegraphics[width=.7\textwidth]{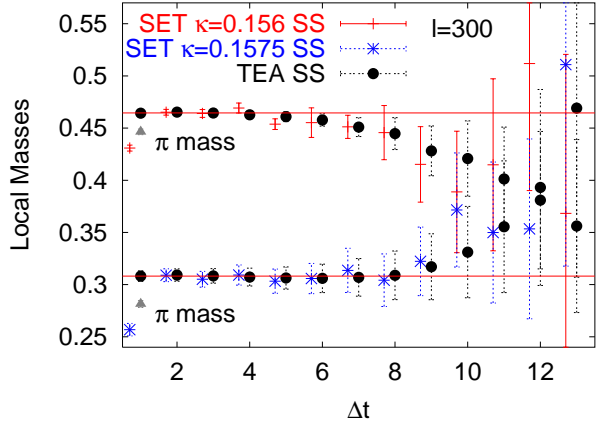}
\end{center}
\caption{ Comparing mass plateaus 
  of the singlet correlator obtained with SET and TEA~\cite{tea}. Smeared
  sources and sinks are used throughout. Notations as in
  Fig.~\ref{Fig:dual_filtering}.}
\label{Fig:dual_filtering_stoch}
\end{figure}
One might still worry about possible systematic errors from spectral
truncation in the singlet mass plateaus contained in \fig{Fig:dual_filtering}.
In this context it is reassuring to observe nice agreement within errors with
the results obtained from stochastic sources, as illustrated in
\fig{Fig:dual_filtering_stoch}.

\begin{figure}[t]
\vskip 1 cm
\hskip -2 cm
\centerline{
\includegraphics[bb = 50 00 309 403, width = .3\textwidth]{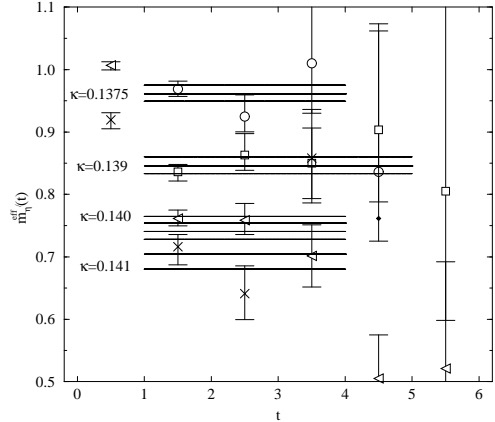}} 
\label{Fig:cppacs}
\hskip -.5 cm
\caption{State-of-the-art mass 
  plateau formation from optimized smearing as obtained in a comprehensive
  study by the CP-PACS group, at their smallest lattice spacing, $a = .1076$
  fm, on a $24^3 \times 48$ lattice, for various sea quark masses. Quotation 
  from Ref.~\cite{cppacs}.}
\end{figure}

It is interesting to compare the plateau formation from dual filtering
(\fig{Fig:dual_filtering}) with the state-of-the-art results from optimized
smearing as obtained in the most recent comprehensive studies of the CP-PACS
collaboration~\cite{cppacs} (\fig{Fig:cppacs}).  The comparison teaches us that
dual filtering is indeed remarkably effective in unravelling the mass plateaus
in the pseudoscalar singlet mesons.

\section{Towards realistic physics results}

\subsection{Unquenched two flavour world}
\label{two_flavour}
Given the good quality of signals within the SESAM setting, we are prepared to
compute the entire mass trajectory of singlet pseudoscalars.  In
table~\ref{tab:simdat} we list the essential run parameters of the SESAM
project, which is based on vacuum field configurations from standard Wilson
fermions on a $16^3\times 32$ lattice~\cite{sesam}, with two active sea quark
flavours, at one value of the coupling, $\beta = 5.6$. Four different sea
quark masses have been used.  The most interesting control parameter is the
ratio between the pion and $\rho$-meson masses as determined on the lattice;
it is quoted in the second column of table~\ref{tab:simdat}.

\begin{table}[b]
\caption{Simulation parameters used at $\beta = 5.6$ and  numbers of
stochastic sources $N_s$ (see Eq.~ \ref{Eq:av_stoch} and
Fig.~ \ref{Fig:dual_filtering_stoch}).
 Last column: numbers of available decorrelated 
vacuum field configurations, $N_{conf}$. }
\label{tab:simdat} 
\begin{center}
\begin{tabular}{|ccccc|}
\hline
$\kappa_{sea}$ & $m_{\pi}/m_{\rho}$ & $L^3*T$ & $N_s$
& $N_{conf}$ \\ 
\hline 
0.1560 & 0.834(3)  & $16^3*32$  & $400$ & $195$ \\
\hline
0.1565 & 0.813(9) & $16^3*32$ & $400$  & $195$ \\
\hline
0.1570 & 0.763(6) & $16^3*32$ & $400$ &  $195$ \\
\hline
0.1575 & 0.692(10) & $16^3*32$ & $400$ &  $195$ \\
\hline
\end{tabular}
\end{center}
\end{table}

\begin{figure}[t]
\begin{center}
\includegraphics[width=.5\textwidth,angle=270]{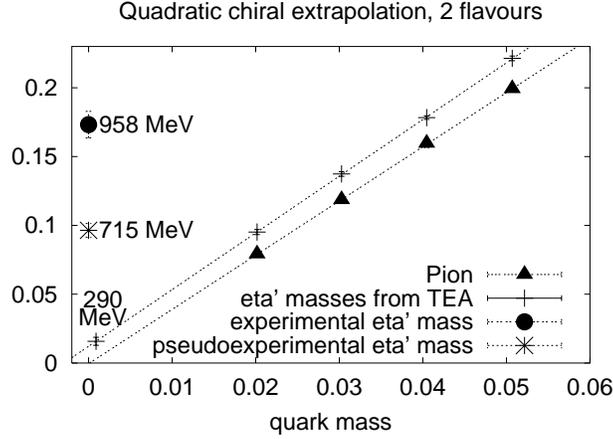}
\end{center}
\caption{ Plot of $m_{\eta}^2$ versus quark mass, all given in lattice
 units~\cite{neff_boston}.}
\label{Fig:trajectory}
\end{figure}

This setting allows for a chiral mass extrapolation of the singlet
pseudoscalar meson composed of $u$ and $d$ quarks, as plotted in
\fig{Fig:trajectory}. The plot contains the mass trajectory for the singlet
channel as well as the nonsinglet extrapolation, marked  by '$\pi$'. We
conclude  that the data allow for a safe chiral extrapolation in both
channels, though there is definite need to lower the sea quark masses in the
simulations.  The precision is sufficient to resolve the mass gap between
singlet and nonsinglet pseudoscalars.

As we have excluded strange quarks altogether, we can of course not expect to hit
the experimental $\eta'$ mass: the extrapolated singlet mass of $290$ MeV
should be compared to the $\eta$ rather than to the $\eta'$ mass.  It is
evident, that we must address the issue of treating strange quarks in addtion
to $u$ and $d$ quarks.

Another shortcoming of SESAM is its restriction to one coupling $\beta$ which
does not allow for a continuum extrapolation.  Very recently, the CP-PACS
collaboration has found in their  $N_f=2$ study with an improved action that this
extrapolation tends to induce a considerable increase in the pseudoscalar
singlet mass~\cite{cppacs}.

\subsection{Partially quenched scenario for the three flavour world}
\label{Sec:partially_quenched}
One of the main restrictions of todays unquenched QCD simulations is their
limitation to two mass degenerate dynamical sea quark flavours.  This presents
a serious shortcoming when dealing with the real physics of the $\eta$-$\eta'$
system.  Nevertheless, the {\it partially quenched approach}~\cite{part_quen}
offers a crutch to include the strange quark sector into the analysis.  In the
partially quenched scenario, strange quarks might appear as valence quarks
only and not contribute at all dynamically to the quark sea.

For the sake of discussion, let us depart from the fully quenched situation.
In this setting, the $s$-loop $s$-loop correlator, $D_{ss}$ in
\eq{Eq:corr_mat}, amounts to the occurence of a double-pole:
$$m_{0,ss}^2/(p^2+m_s^2)^2$$
in momentum space.  In this expression, $m_s$
denotes the valence approximation mass estimate for the pseudoscalar
$\bar{s}s$ meson that can be determined on the lattice from the connected
piece of the $C_{ss}$ correlator.  
As we have seen in section
\ref{Sec:singlet}, the double pole impedes an exponential decay in $t$ and
hence will require a different strategy of analysis: it can
readily be rewritten as a derivative on the single pole expression
\begin{equation}
m_{0,ss}^2/(p^2+m_s^2)^2 = - \frac{\partial}{\partial m_s^2} \;\frac{m_{0,ss}^2}{p^2+m_s^2}\; .
\label{Eq:deriv}
\end{equation}
Iff $m_{0,ss}^2$ is a constant {\it w.r.t}. $p^2$, \eq{Eq:deriv} is readily
Fourier transformed and predicts a linear t-dependence in the ratio of
disconnected to connected correlators to look for:
\begin{equation}
 R(t) := \frac{D_{ss}(t)}{C_{ss}(t)} \stackrel {!}{=}    m_{0,ss}^2 /(2m_s)
 \times t  \; .
\label{Eq:double_pole}
\end{equation}
Note that prior to the exploitation of this ratio relation on actual data,
ground state filtering the $C_{ss}$-data is highly advisable, as explained in
section \ref{scylla}.  The effective strength of the double pole,
$m_{0,ss}^2$, can be extracted from \eq{Eq:double_pole} and is interpreted as
the mass shift due to OZI-rule violation dynamics in channel 2, much in the
spirit of \eq{Eq:gap}:
\begin{equation}
m_{ss}^2 =   m_s^2  + m_{0,ss}^2 \; .
\label{Eq:mass_shift}
\end{equation}
Let us turn now to the
partially quenched setting. We  wish  to  consider
a situation with broken $SU(3)$ flavour
symmetry, where alle valence and sea quarks can differ in masses.

\begin{figure}[t]
\vskip -.8cm
\includegraphics[bb = 00 50 200 300,
width=.22\columnwidth,angle=270]{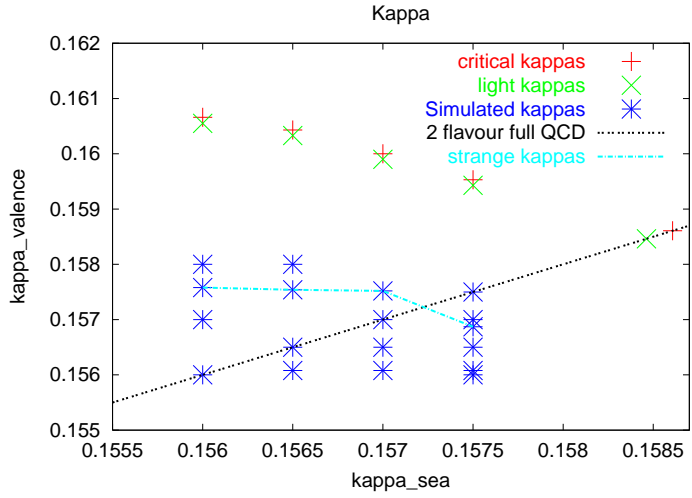}
\vskip 5cm
\caption{Our points of simulation  
  in the hopping parameter plane of valence/sea quarks.}
\label{fig:operational}
\end{figure} 

According to Ref.~\cite{part_quen} partial quenching of a particular quark
species is achieved by introduction of scalar pseudoquark partners (of equal
mass) that freeze those quark degress of freedom inside the determinant. In
this context, one introduces a basis of states with $N_q$ quarks and $k$
additional pseudoquarks.  In our situation, we have $N_q = 5$ (three valence
and two sea quarks) and $k = 3$, if we allow for different masses of sea and
valence quarks.  The correlator has to make reference to these degrees of
freedom, $i,j$.  In Fourier space, it is modeled by the momentum space
ansatz~\cite{kilcup}
\begin{equation}
\label{Eq:partial_quenched}
{\cal C}_{ij} = \frac{\delta_{ij}\epsilon_i}{p^2 + m_i^2} - 
\frac{m_{0}^2}{(p^2+m_i^2)(p^2+m_j^2)F(p^2)}\; \quad , i,j  = 1,2, \dots N_q+k \quad .
\end{equation}
The quantity $\epsilon_i$ is equal to +1 (-1) for quark (pseudoquark =
determinant eater) channels.  The function $F(p^2)$ can be viewed as to model
the implicit multiloop contributions from sea quarks to the disconnected pole
terms:
\begin{equation}
\label{Eq:F-term}
F(p^2) = 1 + \sum_{\mbox{sea quarks } \sigma}\frac{m_{0}^2}{p^2+m_{\sigma}^2}\; .
\end{equation}
$m_i$, $m_j$ and $m_{\sigma}$ denote neutral pseudoscalar nonsinglet meson
masses, while  $m_0$ refers to the effective singlet interaction that
might depend on the quark species involved.

$F$  reads in the case of the SESAM simulation with degenerate dynamical
$u$ and $d$ quarks
\begin{equation}
\label{Eq:F-sesam}
F(p^2) = 1 + \frac{2m_{0}^2}{p^2+m_{\sigma}^2}\; .
\end{equation}
Note that the model incorporates both the fully quenched limit, where $F(t)
\equiv 1$, as well as the consistently $N_f = 2$ unquenched situation, where
$u$ and $d$ valence and dynamical quark masses coincide, such that $m_1 = m_2
= m_{\sigma}$: in this latter instance, the double pole term in the
nonstrange sector is lifted and turned into a displaced single pole term, in
accordance with the considerations  presented in section~\ref{scylla}.

For practical matters, a partially quenched  
lattice analysis 
of Eq.~\ref{Eq:partial_quenched}
is carried out in two steps as
follows
\begin{enumerate}
\item The pseudoscalar meson masses, $m_1, m_2, m_{\sigma}$, are determined
  first by LQCD from connected pseudoscalar correlators with the appropriate
  valence quark settings (in the case of $m_{\sigma}$ and $m_{0,\sigma}$, sea
  and valence quark masses are identified),
%\item the mass gap $m_{0,\sigma}$ is evaluated inside the $u-d$ sector from a
%  two-loop analysis with both valence quark masses equal to the sea quark
%  mass, in very much the same manner as outlined  in section \ref{scylla}. 
\item the singlet correlator matrix contains a set  of
  mass gap parameters, $m_{0,ij}$. They can be 
  computed finally  from a fit of the Fourier transform of the ansatz,
  \eq{Eq:partial_quenched}, to the lattice data for ${\cal C}(t)$.
\end{enumerate}
The latter fitting can be done time-slicewise, yielding fit parameters for
'local' effective mass gaps which should exhibit mass plateau formation.  It
goes without saying that throughout all the numerical analysis steps one
should use the dual filtering method with 'synthetic' data in the sense of
section \ref{scylla}.

\subsection{Partially quenched results}

In a comprehensive study~\cite{ref:tsuk} we analyzed some 800 independent
SESAM configurations on $16^3\times 32$ lattices at $\beta =
5.6$~\cite{ref:sesam}, with four different sea quark masses. In
Fig.\ref{fig:operational} we have plotted the set of hopping parameters of
valence quarks in perspective with their critical values, for our different
sea quark settings.  The full QCD situation with 2 quark flavours in the
sense of chapter \ref{two_flavour} is indicated by the dotted line,
$\kappa_{sea} = \kappa_{valence}$ and the thin line connects the hopping
parameters for strange quarks.

\begin{figure}[h]
\vspace{9pt}
\includegraphics[width=.7\columnwidth,angle=270]{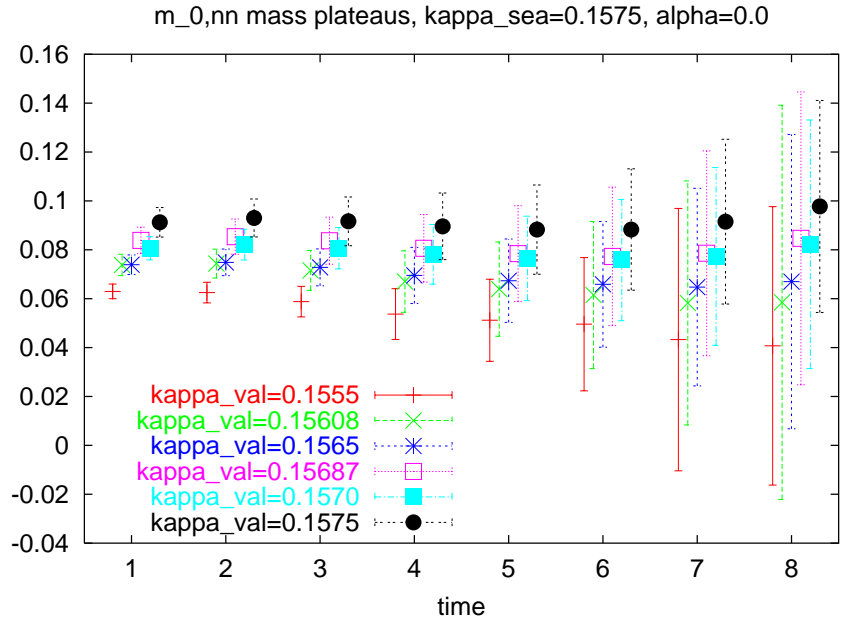}
\caption{First evidence  of plateau formation
  in a partially quenched setting, illustrated on the effective mass gap,
  $m_{0nn}(t)$, in the disconnected two loop correlator, $D_{nn}(t)$. The data refer
  to the lightest sea quark mass of SESAM, i.e. $\kappa_{sea} = .1575$, for
  various isosinglet valence quark masses.}
\label{fig:plateau_signal}
\end{figure}

\begin{figure}[t]
\vspace{9pt}
\includegraphics[width=.7\columnwidth,angle=270]{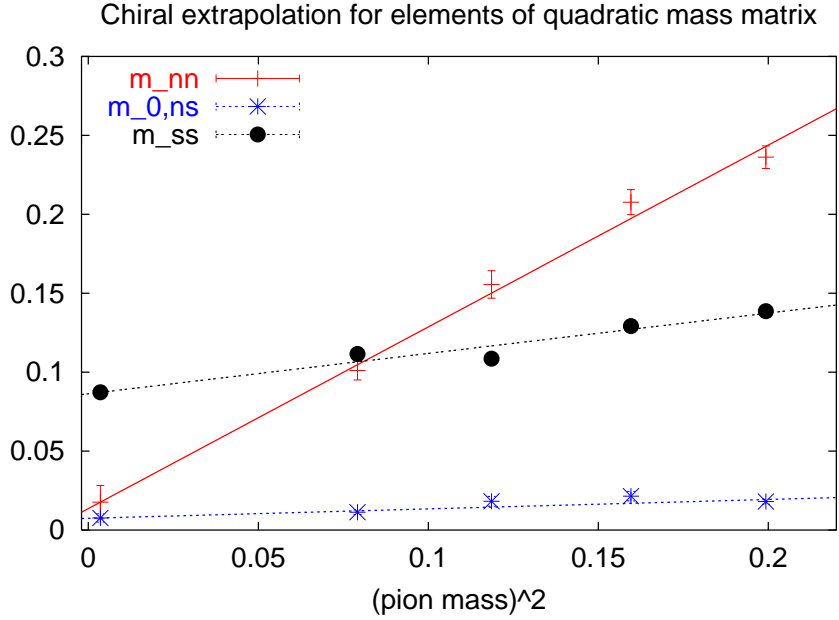}
\caption{Extrapolations of the quadratic mass matrix  elements of light ($n$) and strange ($s$)
 isosinglet  valence quarks   to QCD with {\it chiral} sea quarks (Eq.~\ref{eq:MM}), at $\alpha
 = 0$.}
\label{fig:chiral_signal}
\end{figure}

We analyse the data according to Eq.~\ref{Eq:partial_quenched}.  In doing so
we proceed to study effective mass gaps as determined from the hairpin
diagrams, $D_{ij}(t)$, by fitting, {\it w.r.t.}  $m_{0ij}(t) := \mu$ at each
value of $t$, the zero-momentum Fourier transforms of ansatz
Eq.~\ref{Eq:partial_quenched} to our data.  In broken SU(3) with two active
quark flavours, we introduce two isosinglet pseudoscalars, designated by the
indices $n$ (for `light', nonstrange) and $s$ (for strange). In this manner,
we can study three types of hairpin diagrams, $D_{nn}$ , $D_{ns}$ , $D_{ss}$,
with the strange quark mass determined on the lattice from the kaon mass (for
$\kappa$ locations, see Fig.\ref{fig:operational}).  As a result, we find
satisfying plateau formations in $m_{0ij}(t)$ as illustrated in
Fig.~\ref{fig:plateau_signal}.  In fact, the numerical quality of our signals
comfortably allows for the consecutive extrapolations {\it (i)} to light
valence and then {\it (ii)} to light sea quark masses; the chiral sea quark
extrapolation is exhibited in Fig.~\ref{fig:chiral_signal}.

%xxx

\begin{figure}[h]
\vskip -.4cm
\includegraphics[width=.7\columnwidth,angle=270]{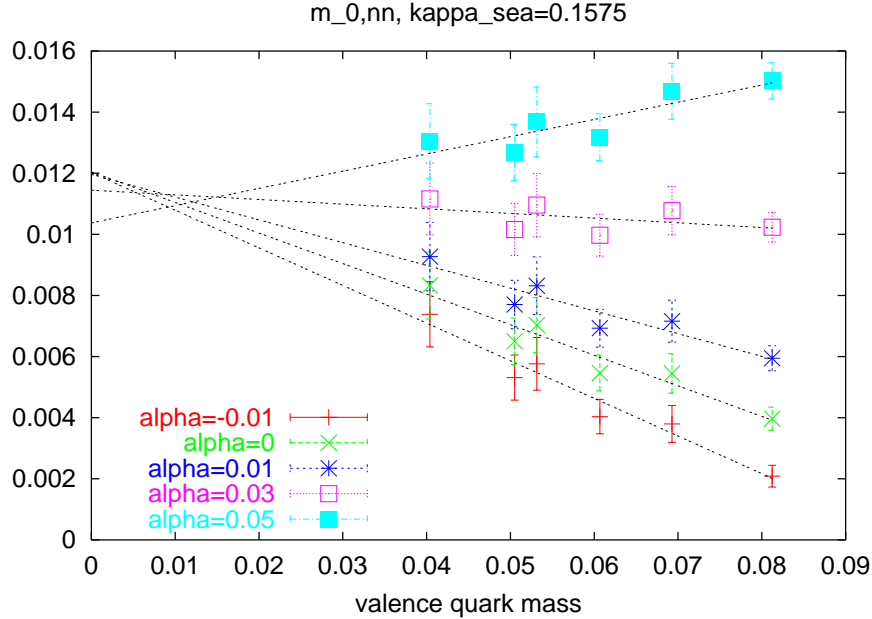}
\caption{Minimizing the  valence quark dependence of the mass gap by 
variation of   $\alpha$.}
\label{fig:minimize}
\end{figure}

From the consecutive chiral extrapolations, we obtain for the
quadratic mass matrix in the quark-flavour basis
\begin{eqnarray}
 {\cal M} := \left( 
\begin{array}{cc}
 m^2_{nn}  &  m^2_{0,ns}\\
 m^2_{0,sn}  &  m^2_{ss} 
\end{array}
\label{eq:MM}
\right) \; ,
\end{eqnarray}
where $m^2_{nn} := M^2_{nn} + m^2_{0nn}$.
With  the $\rho$-meson mass scale~\cite{sesam} we find
\begin{eqnarray}
{\cal M} =
 \left( 
\begin{array}{cc}
 (306 \pm 91)^2 &  \sqrt{2}(201 \pm 34)^2\\
 \sqrt{2}(201 \pm 34)^2  &  (680 \pm 15)^2 
\end{array}  
\right) \mbox{MeV}^2 \; .
\nonumber
\end{eqnarray}
Diagonalization  of $\cal M$ renders 
\begin{equation}
 M_{\eta} = 292 \pm 31 \; \mbox{MeV}, \quad M_{\eta'} = 686 \pm 31 \;
\mbox{MeV}.
\label{eq:values}
\end{equation}
These numbers from $N_f=2$ look promising: the individual mass values --
while being  low {\it w.r.t.} the real three flavour world -- 
yield a  mass splitting that  compares  very well to  phenomenology.

\subsection{Data in accord to  $\chi$PT?}
So far we have neglected that the partially quenched scenario
has been devised in the framework of chiral perturbation
theory ($\chi$PT) where we expect to deal with a limited
number of effective couplings.
In fact,  Bernard and
Golterman~\cite{part_quen} showed that the effective 
chiral symmetry violation (due to  the chiral anomaly)
is  most simply  
described  by
the following contribution to the chiral Lagrangian:
\begin{equation}
{\cal L}_0 =  + \frac{N_f}{2}\Big [{\mu^2} (\eta')^2 +
{ \alpha}
(\partial_{\mu}\eta')^2\Big ] \; ,
\label{Eq:BG}
\end{equation}
{\it i.e.} in terms of just two constants, $\mu$ (mass gap in the chiral
limit) and $\alpha$ (interaction parameter) only!  If the tree approximation
of $\chi$ PT to their partially quenching scenario holds, the simulation data
should be fitted in terms of these two constants, irrespective of the
(supposedly light!!)  valence quark masses chosen.  We will now address the
question whether this is indeed the case. To this end we have to first rewrite
the correlators, Eq.~\ref{Eq:partial_quenched}, by including the parameter
$\alpha$.  Actually, the Lagrangian translates into two-point correlation
functions (propagators) for the neutral pseudoscalar quark bilinears
\begin{equation}
\Phi_{ii}(x) = \bar{q}_i(x)\gamma_5 q_i(x)\; , 
\label{eq:lagr}
\end{equation}
where the index $i$ runs over the quark and pseudoquark {\it
  d.o.f.}~\cite{part_quen}.  

\begin{figure}[h]
\vskip -.3cm
\includegraphics[width=.7\columnwidth,angle=270]{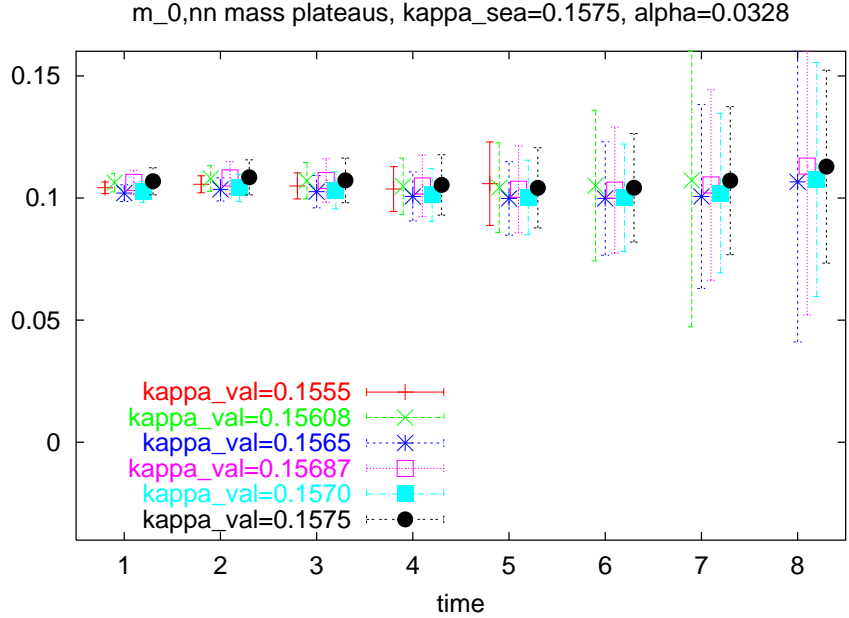}
\caption{The analogue of Fig.~\ref{fig:plateau_signal}: Universal plateau formation, at the lightest sea quark mass,
  when adjusting $\alpha$ to the value $ 0.0328$.}
\label{fig:universal}
\end{figure}
 
In momentum space the disconnected part of these ps-correlators again
has a compact form~\cite{ref:GP} with the infamous double pole structure
\begin{equation}
D_{ij}(p) =  {\cal N} \frac{({ \mu^2} +  \alpha p^2)(p^2+
  M_d^2)}{(p^2+M_{ii}^2)(p^2+M_{jj}^2)(p^2+M_{\eta'}^2)} \; .
\label{eq:form}
\end{equation}
The prefactor, ${\cal N} = 1/(1 + N_f \alpha )$ and the singlet mass, $
M^2_{\eta'} = (M^2_d + N_f \mu^2)/(1 + N_f \alpha )$ are seen to carry
explicit $\alpha$-dependencies. Rememeber that the octet masses
($M_{ii}$,$M_{jj}$, and $M_d$) from valence and sea quarks, respectively, are
readily computed from connected diagrams on the lattice. Our present concern
is the determination of the mass gap, $\mu$, and $\alpha$ from our lattice
data.

We remark that the outcome of the analysis at this stage seems not to be in
accord with the tree approximation to the Lagrangian, Eq.~\ref{eq:lagr}, as
the latter predicts the mass plateaus to be {\it independent} of the valence
quark mass, $m_v$, contrary to the apparant findings in
Fig.~\ref{fig:plateau_signal}.  Therefore, we repeated the above mass plateau
study on a set of nonvanishing $\alpha$-parameters, in an attempt to verify
the compliance of our data with such independence: inspecting the plot of
$m_{0nn}$ vs. $m_v$ (Fig.~\ref{fig:minimize}) it becomes obvious {\it (i)}
that $m_{0nn}(m_v)$ shows a simple, linear behaviour and {\it (ii)} that
$\alpha$ can indeed be adjusted to produce a zero slope of
$m_{0nn}(m_v)$~\footnote{The situation is less favourable for the gap values
  extracted from $D_{ns}$\cite{neff_new}.}!

In this way it is straightforward to find an optimal value, $\alpha_{opt}$,
that eliminates (for a given sea quark mass) the $m_v$-dependence from
$m_{0nn}$, resulting in a universal plateau level.  This collapse of data into
universal plateaus (of heights $\tilde{\mu}$) is exemplified in
Fig.~\ref{fig:universal}.  The comparision with the situation encountered with
$\alpha = 0$ (Fig.~\ref{fig:plateau_signal}) provides clear evidence for our
sensitivity in determining $\alpha_{opt}$.

 After chiral sea quark extrapolation we arrive at a first
estimate of $\alpha$:
\begin{equation}
\alpha = 0.028 \pm 0.013 \quad \mu = 203 \pm 34 \mbox{MeV} \; .
\end{equation}
Note that the numerical value for the mass gap {\it at the chiral point} is
robust {\it w.r.t.}  the two different approaches presented here. One might
phrase it like this: the allowance for a nonvanishing $\alpha$ parameter is
tantamount to the anticipation of the respective chiral valence quark limit
in the mass gap!

%\begin{figure}[htb]
%\vspace{9pt}
%\includegraphics[width=.7\columnwidth,angle=270]{FIGS/alpha_1575.ps}
%\caption{The optimal univerval plateau, at the lightest sea quark mass, with
%$\alpha = 0.0328$.}
%\label{fig:best_alpha}
%\end{figure}

\section{Conclusion}
The application of spectral techniques provides unprecedentedly high accuracy
in the study of hairpin diagrams. Spectral methods thus provide access to
detailed studies of the flavour singlet mesons.  In this work, this has been
demonstrated in two flavour QCD as well as in the $N_f=2$ partially quenched
scenario, with standard Wilson fermions and in the regime of medium mass sea
quarks (see Fig.~\ref{fig:operational}.  Since spectral methods are at their
best in the truly chiral regime, this looks very promising in regard to future
simulations of flavour singlet mesons with realistically light sea quarks,
that we expect to come up within the overlap fermion formulation on the
lattice.

\end{document}